\documentclass[twocolumn]{aastex61}
\usepackage{graphics,graphicx,natbib,subfigure,wrapfig}
\usepackage{ textcomp }
\usepackage{amssymb}%,deluxetable}

\newcommand{\ee}[1]{\mbox{${} \times 10^{#1}$}}% scientific number format
\newcommand{\eten}[1]{\mbox{$10^{#1}$}}% power of ten
\newcommand{\msun}{\mbox{M$_\odot$}}% Msun
\newcommand{\water}{H$_2$O}

\newcommand{\cotwo}{CO$_{2}$}

\newcommand{\hh}{\mbox{{\rm H}$_2$}}
\newcommand{\nn}{\mbox{{\rm N}$_2$}}
\newcommand{\ammonia}{\mbox{{\rm NH}$_3$}}

\newcommand{\methanol}{\mbox{{\rm CH$_3$OH}}}

\begin{document}

\title{Unlocking CO Depletion in Protoplanetary Disks II. Primordial C/H Predictions Inside the CO Snowline}

\correspondingauthor{Kamber R. Schwarz}
\email{kschwarz@lpl.arizona.edu}

\author{Kamber R. Schwarz}
\altaffiliation{Sagan Fellow}
\affiliation{Department of Astronomy, University of Michigan, 1085 South University Ave., Ann Arbor, MI 48109, USA}
\affiliation{Lunar and Planetary Laboratory, University of Arizona, 1629 E. University Blvd, Tucson, AZ 85721, USA}
\author{Edwin A. Bergin}
\affiliation{Department of Astronomy, University of Michigan, 1085 South University Ave., Ann Arbor, MI 48109, USA}
\author{L. Ilsedore Cleeves}
\affiliation{Department of Astronomy, University of Virginia, 530 McCormick Rd, Charlottesville, VA 22904},
\author{Ke Zhang}
\altaffiliation{Hubble Fellow}
\affiliation{Department of Astronomy, University of Michigan, 1085 South University Ave., Ann Arbor, MI 48109, USA}
\author{Karin I. \"{O}berg}
\affiliation{Harvard-Smithsonian Center for Astrophysics, 60 Garden Street, Cambridge, MA 02138, SA},
\author{Geoffrey A. Blake}
\affiliation{Division of Geological \& Planetary Sciences, MC 150-21, California Institute of Technology, 1200 E California Blvd, Pasadena, CA 91125}
\author{Dana E. Anderson}
\affiliation{Division of Geological \& Planetary Sciences, MC 150-21, California Institute of Technology, 1200 E California Blvd, Pasadena, CA 91125}

\begin{abstract}
CO is thought to be the main reservoir of volatile carbon in protoplanetary disks, and thus the primary initial source of carbon in the atmospheres of forming giant planets. However, recent observations of protoplanetary disks point towards low volatile carbon abundances in many systems, including at radii interior to the CO snowline. One potential explanation is that gas phase carbon is chemically reprocessed into less volatile species, which are frozen on dust grain surfaces as ice. This mechanism has the potential to change the primordial C/H ratio in the gas. However, current observations primarily probe the upper layers of the disk. It is not clear if the low volatile carbon abundances extend to the midplane, where planets form.
We have run a grid of 198 chemical models, exploring how the chemical reprocessing of CO depends on disk mass, dust grain size distribution, temperature, cosmic ray and X-ray ionization rate, and initial water abundance. 
Building on our previous work focusing on the warm molecular layer, here we analyze the results for our grid of models in the disk midplane at 12~au. We find that either an ISM level cosmic ray ionization rate or the presence of UV photons due to a low dust surface density are needed to chemically reduce the midplane CO gas abundance by at least an order of magnitude within 1~Myr. In the majority of our models CO does not undergo substantial reprocessing by in situ chemistry and there is little change in the gas phase C/H and C/O ratios over the lifetime of the typical disk.
 However, in the small sub-set of disks where the disk midplane is subject to a source of ionization or photolysis, the gas phase C/O ratio increases by up to nearly 9 orders of magnitude due to conversion of CO into volatile hydrocarbons.
\end{abstract}


\section{Introduction}
Planets are formed out of the gas and dust in the protoplanetary disks around young stars. The composition of these planets is thus primarily set by the composition of the parent disk. A planet formed via gravitational instability should have an atmospheric composition similar to that of the bulk composition of the disk, i.e., that of the host star. For planets formed via core accretion, the connection between disk and planet compositions is more complex. 

A factor in determining a planet's composition in the core accretion scenario is its formation location relative to the snowlines of major volatiles \citep{Oberg11c}.
A snowline is the location in a disk where a species such as CO, \cotwo, or \water\ transitions from being frozen out as ice to being in the gas phase. These condensation fronts result in sharp transitions in the C/O ratio in both the gas and the solids.
The C/O ratio in the gas and solids can be further modified by disk dynamics. As dusty particles drift inward they can remove volatile ices from the outer disk \citep{Ciesla06}. At small radii these ices will sublimate, enriching the inner disk gas in volatiles \citep{Oberg16,Estrada16}. 
Disk evolution can also change the snowline locations due to either radial drift of dust \citep{Piso15} or an evolving temperature structure \citep{Eistrup18}. 
Additionally, chemical reprocessing can change the relative abundance of volatiles in the gas and ice \citep{Yu16,Eistrup16}.

Models combining various aspects of this evolution: disk chemistry, planet migration, and atmospheric chemistry, point toward the use of the C/O ratio in a planet's atmosphere as a way to trace the formation history of the planet when combined with additional information such as the C/H or C/N ratio \citep{Cridland17b,Booth17}. Several studies have now derived C/O ratios for giant exoplanets, either directly from the detection of spectra in the atmosphere or through a combination of observations and models \citep{Kreidberg15,Lavie17,Espinoza17}. As observations and models continue to improve we are fast approaching an era of connecting protoplanetary disk and planet compositions. 

Observations of a growing number of protoplanetary disks reveal low CO abundances relative to that expected from the dust mass \citep{Ansdell16,Long17}. Models of CO in these disks show that this discrepancy cannot be fully explained by CO freeze-out, nor is it due to a failure to properly correct for isotopologue selective self shielding \citep{Williams14,Miotello14}. 
CO is not the only volatile molecule with low observed abundances. Observations of \water\ vapor and atomic carbon also reveal these species to be under-abundant \citep{Kama16,Du17}. This `missing volatiles problem' has several potential solutions including gas disk dispersal, gaseous interactions with the evolving dust population, and chemical reprocessing \citep{Reboussin15,Bai16,Krijt16,Xu17}. However, as disk gas masses are usually derived from either CO or dust observations, it is often difficult to distinguish between the different scenarios. 

One way to differentiate between disk dispersal and mechanisms which affect only the volatiles is through observations of the \hh\ isotopologue HD, which is more closely related to the total gas mass than either CO or dust and primarily emits from warm (20-50 K) gas within the inner few tens of~au \citep{Zhang17}. HD has been successfully detected in three disks to date \citep{Bergin13,McClure16}, and reveals CO to be under-abundant by roughly two orders of magnitude in TW Hya \citep{Favre13,Schwarz16}. For DM Tau and GM Aur, the other disks with HD detections, CO appears under-abundant by an order of magnitude, though uncertainties related to the disk thermal structure remain \citep{McClure16}. These lines of evidence point towards processes beyond gas disk dispersal contributing to low CO-to-dust ratios.

Millimeter observations of CO almost exclusively probe regions in the disk above the midplane and outside the CO snowline, in the warm molecular layer. In only one system have optically thin CO isotopologues been observed inside the CO snowline: TW Hya. Using observations of optically thin $\mathrm{^{13}}$C$\mathrm{^{18}}$O emission interior to the midplane CO snowline in TW Hya \citet{Zhang17} find an average $\mathrm{^{13}}$C$\mathrm{^{18}}$O  abundance of 1.7\ee{-10} relative to \hh\ for gas warmer than 20~K in the radial range 5-20~au. This corresponds to an average CO abundance in the same region of 6.5\ee{-6}. As CO is expected to be the dominant gas phase carbon species between the CO and \cotwo\ snowlines, this suggests that the midplane C/H gas ratio is well below expectations in this one system. 
Previous studies demonstrate substantial chemical reprocessing of CO in the midplane is possible, particularly in the presence of cosmic rays \citep{Yu16,Eistrup16}. 
In this work we explore the viability of chemical reprocessing as a way to remove volatile species from the gas for models spanning a large range of physical conditions, with a focus on different disk masses and large grain fractions. 

In \citet{Schwarz18}, hereafter Paper I, we analyzed the results of our grid of chemical models for the warm molecular layer. In this paper we focus on the midplane CO gas abundance at 12~au, which is between the midplane CO and \cotwo\ snowlines for all models, as well as within the expected formation region for giant planets \citep{Chabrier14}. 
\S \ref{midmodel} briefly summaries our model framework and parameter space. The results are described in \S \ref{midresults}. In \S \ref{middiscussion} we compare our results to previous studies and solar system bodies, as well as discuss the implications for planet formation. Finally, our findings are summarized in \S \ref{midsummary}.

\section{Model}\label{midmodel}
We use the model described in detail in Paper I, summarized below. We explore a range of parameters: disk mass, dust grain size distribution, temperature, X-ray and cosmic ray ionization, and initial water abundance. 
Our model setup is a two-dimensional, azimuthally symmetric disk. The density and temperature structure, as well as the dust opacity, are generated using the radiative transfer code TORUS \citep{Harries00}. We consider disks with an inner radius of 0.1~au and an outer radius of 200~au, and masses of 0.1, 0.03, and 0.003~\msun. Our choice of outer disk radius is larger than the radius of a `typical' protoplanetary disk. However, the midplane chemistry in our models is not sensitive to the outer radius of the disk (see \S \ref{app:smalldisk}). This is due to several factors: the only source of radiation in our models is the central star, we do not physically evolve the disk, and the chemical evolution at each radius is calculated independently

Each disk has two dust populations. The first treats small grains ($r_d = 0.005 - 1 \micron$), which are well mixed with the gas. Large grains ($r_d = 0.005 - 1000 \micron$) are more settled than the small grains and gas as described in Paper I. 
Both the small and large grain populations have an MRN size distribution \citep{Mathis77}.  The fractional dust mass in large grains varies from 0 (all dust in small grains) to 0.99 in eleven steps. 

All of our disk models are irradiated by a central T Tauri star with a mass of 0.8~\msun\ and an effective temperature of 4300 K. The radiative transport of the UV and X-ray photons through the disk are computed using the methods described by \citet{Bethell11a,Bethell11b}. 
Our chemical evolution model is based on that of \citet{Cleeves14} and the chemical networks of \citet{Smith04}, \citet{Fogel11}, and \citet{McElroy13}. The network includes an extensive number of gas phase reactions, including ionization by cosmic rays and X-rays, as well as a limited number of grain surface reactions focusing on the grain surface formation of \water\ and \cotwo\ as well as the hydrogenation of volatile carbon and nitrogen. The initial abundances are listed in Table~\ref{midabun} and are based on the model molecular cloud abundances of \citet{Aikawa99}. The exceptions are the initial \water\ ice and \cotwo\ ice abundances, which have been adjusted to match the carbon and oxygen partitioning assumed by \citet{Oberg11c} in order to facilitate comparison with their C/O values.
We evolve the chemistry for 6~Myr while the disk physical conditions remain static

Our fiducial models assume a cosmic ray ionization rate of 1.6\ee{-19} s$^{-1}$ and an X-ray luminosity of \eten{30} erg s$^{-1}$ for three disk masses and eleven different large grain fractions. 
This lower cosmic ray ionization rate replicates the modulation of cosmic rays by winds \citep{Cleeves14b}.
We also consider high X-ray models (\eten{31} erg s$^{-1}$), high cosmic ray rate models (2\ee{-17} s$^{-1}$) equivalent to the flux in the diffuse ISM, warm models where the gas and dust temperature has been uniformly enhanced by 20 K, even in the midplane, and warm high cosmic ray rate models for a total of 165 different models. Additionally, we remove the initial water abundance for the fiducial, high X-ray, and high cosmic ray models with a disk mass of 0.03~\msun, for a total of 198 unique models.

\begin{deluxetable}{ll}
\tablewidth{0pt}
\tablecaption{Physical model properties}\label{grid}
\tablehead{
\colhead{Parameter} & \colhead{Values}  \\
}
\startdata
M$_{disk}$ ($\msun$) & 0.1, 0.03, 0.003 \\ 
L$_{XR}$ (erg s$^{-1}$) & 1E30, 1E31 \\ 
$\zeta_{CR}$ (s$^{-1}$) & 1.6E-19, 2E-17 \\
$f_l$ & 0.0, 0.1, 0.2, 0.3, 0.4, 0.5,\\
 & 0.6, 0.7, 0.8, 0.9, 0.99\\
R$_{in}$ (au) & 0.1 \\
R$_{out}$ (au) & 200 \\
\enddata
\end{deluxetable}

\begin{deluxetable}{llll}
\tablewidth{0.5\textwidth}
\tablecaption{Initial abundances relative to \hh}\label{midabun}
\tablehead{
\colhead{Species} & \colhead{Abundance} & \colhead{Species} & \colhead{Abundance}\\
}
\startdata
\hh\         & 1.00E00 & H$_3^+$   &      2.00E-08  \\
He          &2.80E-01 & HCO$^+$  &      1.80E-08   \\
CO         & 2.00E-04 & C$_2$H   &      1.60E-08 \\
\water(gr)  &  1.20E-04 & H$_2$CO & 1.60E-08 \\
N         &  4.50E-05 & SO & 1.00E-08 	 \\
\cotwo(gr) & 4.00E-05 & CS    &      8.00E-09	\\  
\nn\    &      2.00E-06 & C$^+$    &      2.00E-09  \\
C     &      1.40E-06  &  Si$^+$   &      2.00E-11	\\
\ammonia  &   1.60E-07 & S$^+$    &      2.00E-11\\
CN    &      1.20E-07  & Mg$^+$   &      2.00E-11 \\
HCN   &      4.00E-08 & Fe$^+$   &      2.00E-11 \\
\enddata
\end{deluxetable}

\section{Results}\label{midresults}
We choose to focus on a single location in the disk, the midplane at 12~au, in order to make the analysis of a large number of models more manageable. We assume that 12~au is representative of the disk midplane between the CO and \cotwo\ snowlines. The validity of this assumption is discussed at the end of this section.

\begin{figure}[!]
\setlength{\intextsep}{0pt}
\centering
    \includegraphics[width=0.5\textwidth]{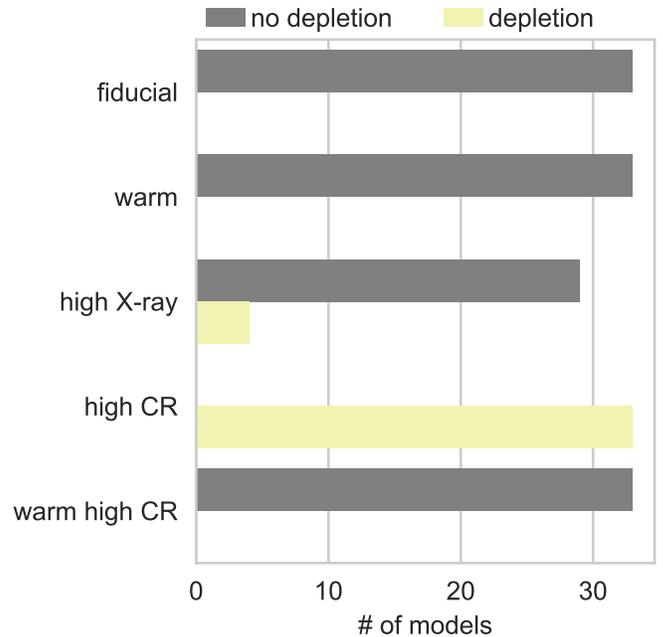}
  \caption{Breakdown of the number of models that are depleted (X(CO gas) $<$ \eten{-4}) and not depleted (X(CO gas) $>$ \eten{-4}) at 1~Myr. Twelve of the high CR models are extremely depleted (X(CO gas) $<$ \eten{-5}). \label{midsumbar}}
\end{figure}

\begin{figure*}[!]
\setlength{\intextsep}{0pt}
    \includegraphics[width=1.0\textwidth]{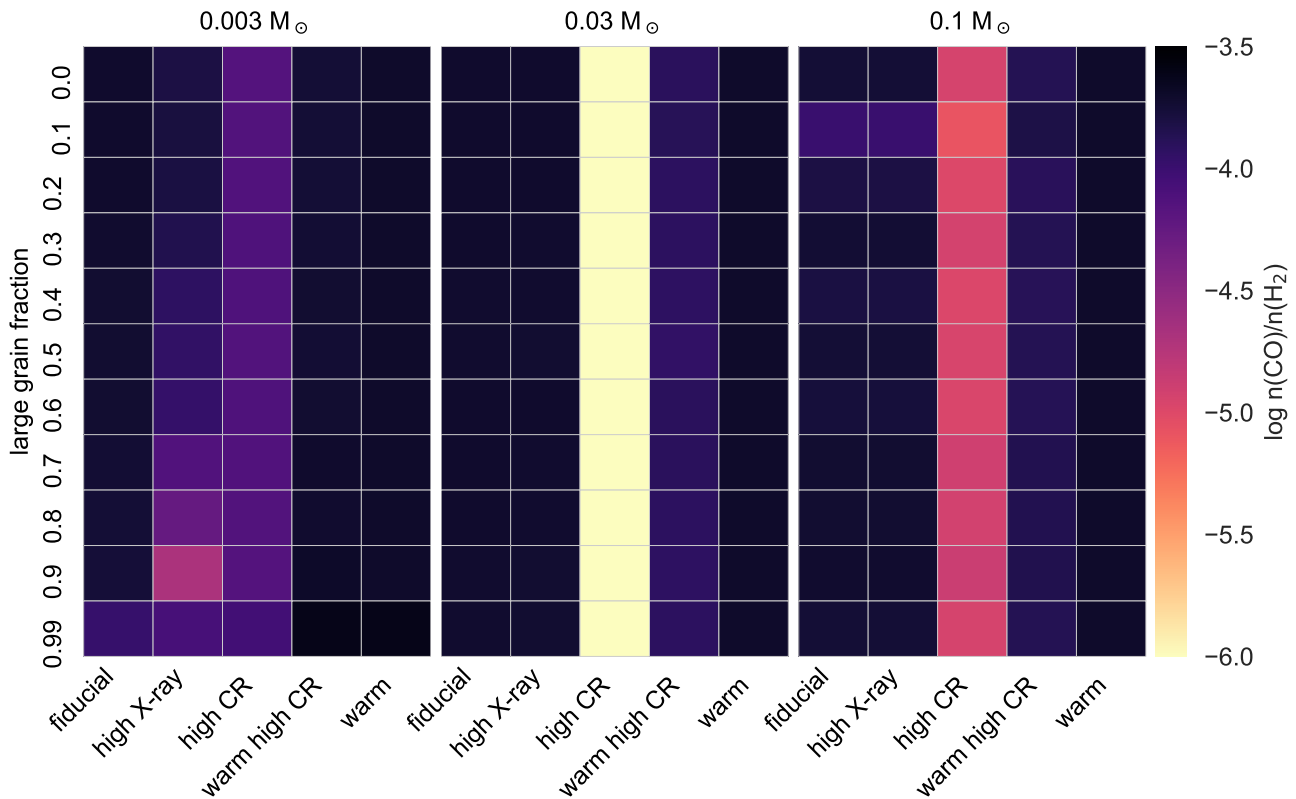}
\caption{Log CO gas abundance relative to \hh\ in the midplane at 12~au for each model after 1~Myr. \label{midheatin}}
\end{figure*}

Figure \ref{midsumbar} summaries our findings for models with a standard ISM abundance.
A model is considered depleted if the gas phase CO abundance relative to \hh\ is less than half the initial abundance, that is $X(CO) < \eten{-4}$. A model is considered extremely depleted when the gas phase CO abundance drops below \eten{-5}, or 20 times lower than the initial abundance.
At 12~au after 0.1~Myr, 1\% of models are depleted by a factor of 2 to 20 in gas phase CO, increasing to 15\% after 1~Myr, 11\% after 3~Myr and 10\% after 6~Myr. Additionally, 7\% of models are depleted by more than a factor of 20 after 1~Myr, increasing to 30\% after 3~Myr and 39\% after 6~Myr.
The majority of the models are not depleted in midplane CO after 1~Myr, with $X(CO) > \eten{-4}$. The models which are able to reduce the CO gas abundance by more than two orders of magnitude within 1~Myr are the 0.03~\msun\ disks with a high cosmic ray ionization rate, while the 0.1~\msun\ high cosmic ray models are depleted by an order of magnitude after 1~Myr. 
The midplane CO abundance after 1~Myr in each model is shown in Figure~\ref{midheatin}. Additionally, the abundances of the top five carbon bearing species in each model are given in Section~\ref{app:table}.

In the high cosmic ray rate models, the carbon has been chemically reprocessed into \methanol\ ice. Cosmic rays are able to create H$_{3}^+$, which reacts with gas phase CO to form HCO$^+$. This almost immediately recombines with an electron, placing the carbon once again in CO. However, this process is also a way to free hydrogen atoms from \hh. Some of these hydrogen atoms freeze out onto grains where they are able to hydrogenate CO ice before it can be thermally desorbed back into the gas. Successive hydrogenation on the grain surface ultimately culminates in the formation of \methanol\ ice. 
Because the process of converting CO gas into \methanol\ ice does not require any additional oxygen beyond that in CO, it is still effective in the models where the initial water is removed. Thus, the CO abundances in the fiducial, high X-ray, and high cosmic ray models without an initial reservoir or water ice reflect the abundances in the corresponding models with a fiducial water abundance. 
\methanol\ ice is an end state product in our chemical network. 
Additional processing on the grain surface is possible, leading to the formation of, e.g., hydrocarbon ices on timescales of several~Myr \citep{Bosman18}. 
As such the \methanol\ ice abundances in our models should be considered upper limits while the overall complexity of the chemistry should be considered a lower limit since \methanol\
 can be a precursor to more complex species.

For the 0.003~\msun\ disk, the fiducial, high X-ray, and high cosmic ray models with 99\% of their mass in large grains are depleted in CO by a factor of two after 1~Myr. In these models carbon is primarily in \cotwo\ ice. 
In these low density, highly settled disks, UV photons are able to reach the midplane. These photons dissociate \water\ ice, creating OH gas. The OH then freezes back onto the grain where it reacts with CO ice to form \cotwo\ ice. As in higher mass disks, the high cosmic ray models also allow for additional chemical reprocessing of CO. 
Likewise, the ionization provided by the X-rays in the high X-ray models also leads to approximately half of the CO being reprocessed. However, the higher temperatures in these models reduce the amount of CO on the grain surface, resulting in longer timescales for the reprocessing into \cotwo\ and \methanol\ ice. A similar behavior is seen in all of the models with both artificially increased temperature and high cosmic ray rates.

\subsection{Radial Abundance Variations}
\begin{figure*}[!]
\setlength{\intextsep}{0pt}
    \includegraphics[width=1.0\textwidth]{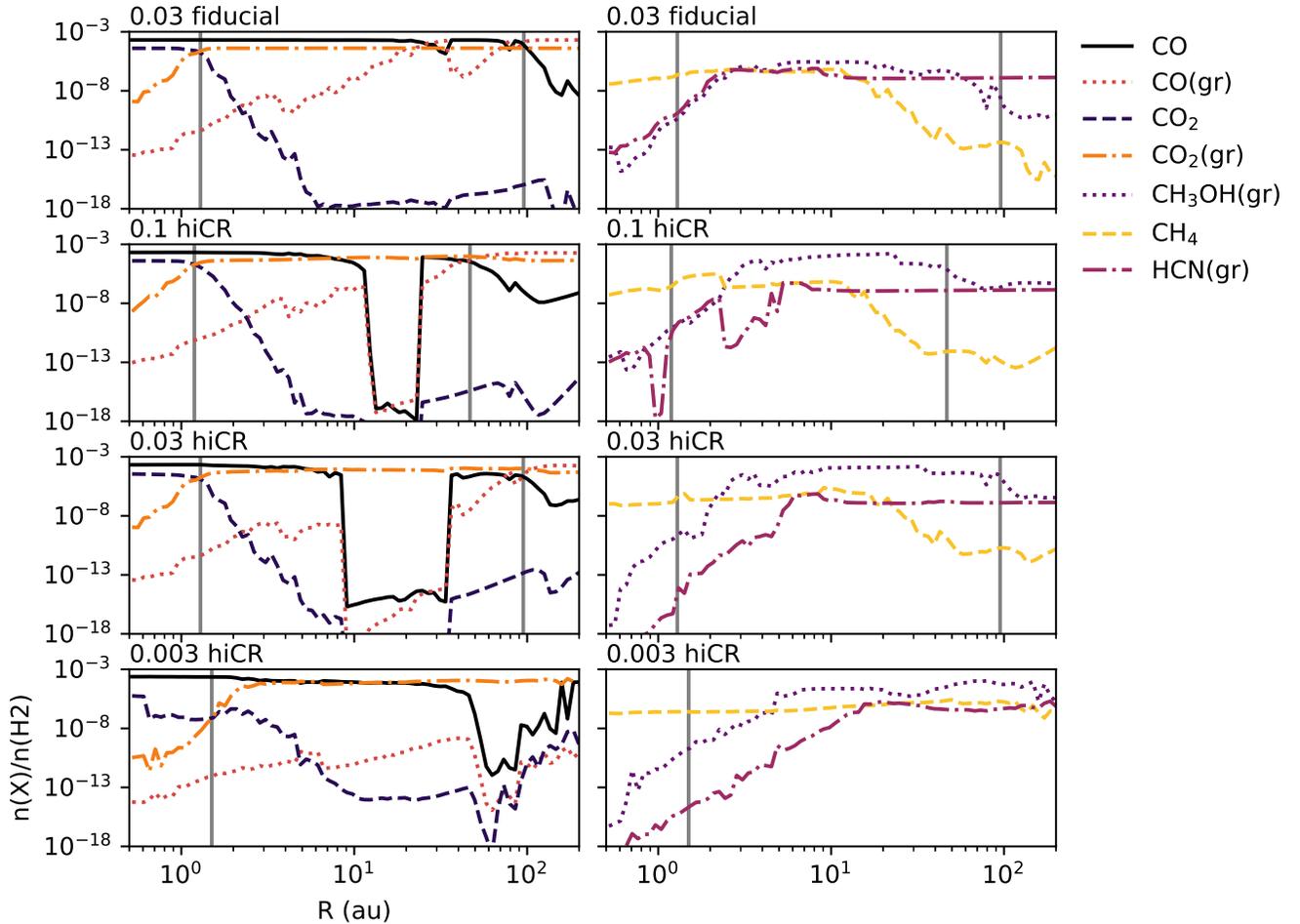}
\caption{Midplane abundances of carbon bearing species as a function of radius for four representative models after 1~Myr. All models have 50\% of their dust mass in large grains. The \cotwo\ and CO snowlines, defined as the largest radius where the gas and ice abundance of the species are equal, are shown by vertical grey lines. For the 0.003~\msun\ disk the entirety of the disk is inside the CO snowline. \label{fullmod1}}
\end{figure*}

\begin{figure*}[!]
\setlength{\intextsep}{0pt}
    \includegraphics[width=1.0\textwidth]{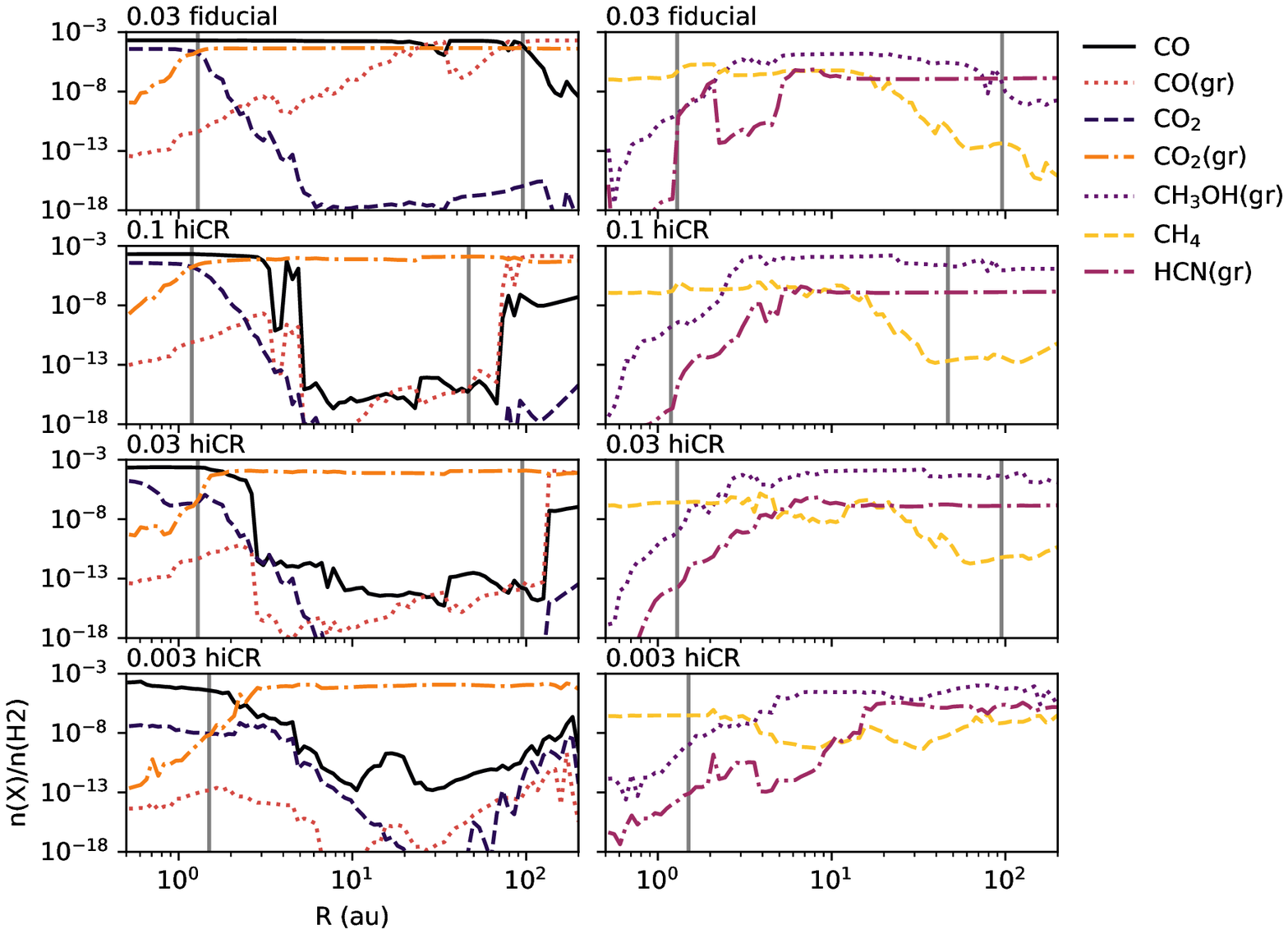}
\caption{Midplane abundances of carbon bearing species as a function of radius for four representative models after 6~Myr. All models have 50\% of their dust mass in large grains. The \cotwo\ and CO snowlines, defined as the largest radius where the gas and ice abundance of the species are equal, are shown by vertical grey lines. For the 0.003~\msun\ disk the entirety of the disk is inside the CO snowline. \label{fullmod6}}
\end{figure*}

To test our assumption that the midplane abundances at 12~au are representative of the disk between the CO and \cotwo\ snowlines we analyze the variation in the radial abundance strucuter of carbon bearing species in the midplane for four representative disks: the fiducial 0.03~\msun\ disk model with 50\% large grains, which does not show evidence for substantial CO reprocessing, as well as the high cosmic ray rate, 50\% large grain models for all three disk masses. The midplane abundance of the major carbon bearing species in these models are shown after 1 Myr in Figure~\ref{fullmod1} and after 6 Myr in Figure~\ref{fullmod6}. In each of these models 12~au is located between the CO and \cotwo\ snowlines.

The timescales for CO reprocessing are shorter in the outer disk, such that after 1~Myr gas phase CO depletion has occurred over a limited radial range. 12~au is within this range for the 0.1 and 0.03~\msun\ disks with a high cosmic ray rate, while in the lower density 0.003~\msun model significant CO reprocessing has occurred only at larger radii.
After 6~Myr, the CO abundance between the \cotwo\ and CO snowlines is more uniform, though the CO gas abundance does increase at the snowlines of carbon bearing ices such as \methanol\ and HCN. In each model the change in CO abundance from the CO snowline to just outside the \cotwo\ snowline is minor enough to not affect whether a disk is classified as depleted in CO according to our criteria. Therefore we conclude that the abundances at 12~au are representative of the disk between these two major snowlines, with the caveat that depletion timescales will be longer at smaller radii. 

\section{Discussion}\label{middiscussion}
\subsection{Chemistry as a Depletion Mechanism}
In our models, there are two sets of conditions which lead to substantial chemical reprocessing of gas phase CO in the inner disk midplane. The first is if the disk has a low enough mass surface density of small grains or overall gas+dust physical density for UV radiation to reach the midplane, reprocessing the carbon primarily into \cotwo\ ice. The second, and more widespread, condition is to expose the disk to a high ionization rate, comparable to the cosmic ray ionization rate seen in the ISM, in which case much of the carbon is placed into \methanol\ ice. 

Our results, namely that ionization is an important mechanism and that carbon goes into \cotwo\ and \methanol\ ices, agree with previous studies of chemical reprocessing \citep{Eistrup16,Yu16}. These studies also included an ISM-level a cosmic ray ionization rate. However, it is possible that Class II disks see a lower cosmic ray rate due to modulation by a stellar wind \citep{Cleeves14}. This modulation of cosmic rays is seen in our own Solar System, while observations of ionized molecules in TW Hya also indicate a low cosmic ray ionization rate \citep{Cleeves15}.
An alternative ionization source is stellar X-rays, which can also result in chemical reprocessing of CO for a range of physical conditions \citep{Dodson18}
However, the ionization structure in most systems remains largely unconstrained by observations. Analysis of the N$_2$H$^+$ and HCO$^+$ emission in these systems, similar to the work by \citet{Cleeves15} in TW Hya, will provide crucial observational constraints on the typical ionization level in protoplanetary disks.

It is possible that depletion primarily occurs in younger Class 0 and Class I systems, which may have a higher cosmic ray ionization rate \citep{Padovani16}. In fact, there is emerging observational evidence of cosmic ray acceleration in the bow shock region of low mass protostars \citep{Tychoniec17}.
Observations of the envelopes of some Class 0 protostars reveal reduced CO gas abundances, suggesting CO depletion may begin early \citep{Anderl16}. Conversely, there is no evidence of substantial CO depletion in the embedded Class I disk L1527 \citep{vantHoff18}.
 Until we understand more about the ionization structure in protoplanetary disks as a population it will be difficult to determine the extent to which chemical reprocessing contributes to the removal of volatiles from the gas.

Dynamical processes could also contribute to CO depletion. Vertical mixing brings gas phase species from the upper layers of the disk to the cold midplane, where they freeze out onto dust grains, thus depleting the upper layers of the disk in what is sometimes referred to as the vertical ``cold finger" effect \citep{Meijerink09}. This freeze out of gas can result in a factor of 50 depletion in the upper layers, while enriching the midplane ices in volatiles \citep{Xu17}. As these icy grains drift inward they enrich the gas inside a given species' snowline \citep{Krijt18}. This will counter the depletion due to chemical reprocessing in the inner disk so long as grains continue to drift inward. Indeed, \citet{Booth17} find that the inward drift of icy pebbles can lead to the formation of giant planets with both super solar C/H and super solar C/O exterior to the \water\ snowline. 
If the dust grains are prevented from drifting inward, e.g., due to a pressure bump, this volatile enrichment will not be observed. Alternatively, if, as in our models, CO ice is converted into species with higher binding energies there will not be an enrichment of gas phase carbon as grains pass inside the CO snowline. However, so long as ice coated grains continue to drift inward the ices will eventually sublimate, resulting in greater enrichment at smaller radii.

\subsection{Additional Factors}
Our models focus on how chemistry in disks with a range of masses and grain size distributions is effected by limiting cases for the cosmic ray ionization rate and X-ray luminosity of T-Tauri stars. We also assume that the chemistry is evolving in a physically static system. In reality, over the course of the 6 Myr we consider the disk, and the central star, will be evolving, leading to changes in the dust size distribution, dust spatial distribution, and stellar luminosity. The luminosity of the central star, particularly in the FUV, can potentially impact the disk chemistry in two ways. First, FUV radiation heats the gas disk through the photoelectric effect \citep{Weingartner01}. The star cools with time and it's UV luminosity decreases, resulting in a cooler disk. 
As the disk cools the snow lines of major volatiles move inward, with the largest change seen for snow lines which start at large radii \citep{Yu16,Eistrup18}.
Our models show that warmer disk temperatures, such as expected for younger disks, hinder chemical reprocessing (Figure~\ref{midheatin}), a result also found by several previous studies \citep{Reboussin15,Bosman18}. 
However, as the star evolves, the decrease in FUV luminosity also means the flux of ionizing photons in the disk decreases. 
This will make it more difficult for CO to be reprocessed into \cotwo\ ice, as this process relies on the photodissociation of \water\ ice. 
So while a cooling disk will promote the reprocessing of CO, a decreasing UV flux will hinder it. Additional work exploring how these two effects work in concert is needed to determine which one is dominant.

There is currently some debate in regards to whether disks inherent their initial abundances from the ISM or if the chemistry is at least partially reset by heating during collapse. However, after several million years of chemical evolution, models with atomic initial abundances have a very similar composition to models with molecular initial abundances \citep{Eistrup18,Molyarova17}.
More generally, a reduction in the initial amount of \water\ ice will slow down the conversion of CO to \cotwo\ ice. To form \cotwo, CO reacts with OH on the grain surface and OH is formed from dissociated \water.

\subsection{Consequences for Planet Composition}
\subsubsection{C/H}
\begin{figure}
\setlength{\intextsep}{0pt}
\centering
    \includegraphics[width=0.5\textwidth]{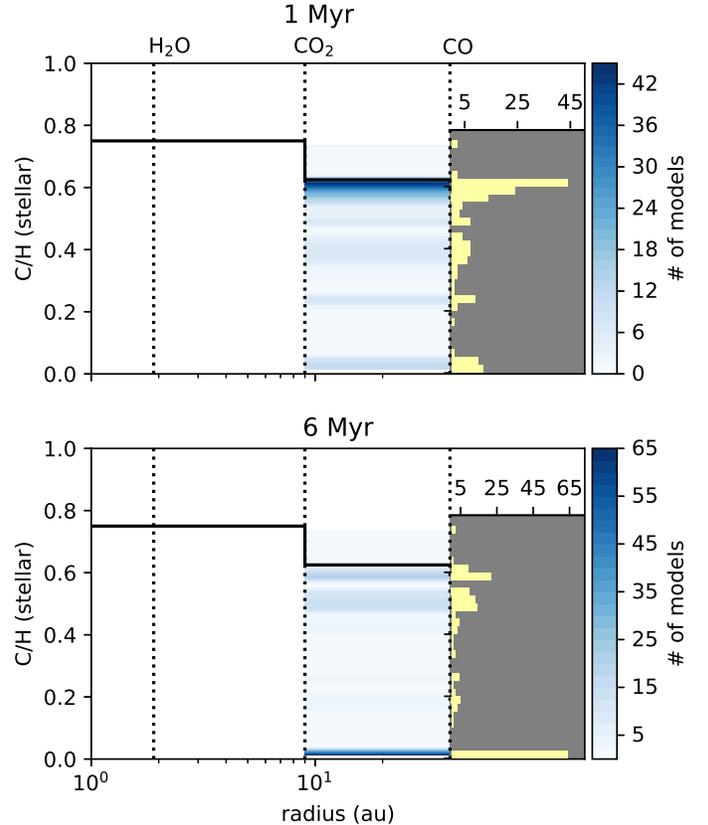}
\caption{C/H ratio in the gas based on the location of major volatile snowlines (line) and the distribution of C/H values inside the CO snowline for our models (shading \& histogram) after 1~Myr (top) and 6~Myr (bottom). Radial changes in the C/H value are based on the carbon partitioning assumptions of \citet{Oberg11c}.
\label{chplot}}
\end{figure}

In this section we analyze the total gas phase C/H ratio in the midplane at 12~au to assess how much bulk gas phase carbon planets can accrete in their atmospheres.
Figure \ref{chplot} shows the C/H ratio in the gas considering only the snowline locations for the major volatiles, assuming 62.5\% of the total carbon content is in CO, 12.5\% is in \cotwo, and the remaining 25\% is in refractory material such as carbon grains. 
Between the CO and \cotwo\ snowlines all of the gas phase carbon is in CO, so the gas phase C/H ratio is 0.625 relative to stellar. Inside the \cotwo\ snowline \cotwo\ returns to the gas and the gas C/H ratio rises to 0.75.

Also shown in Figure~\ref{chplot} is the distribution of gas phase C/H values in the midplane at 12~au for our models. In this framework, the largest possible C/H value in our models, relative to stellar, is 0.75.
Between the CO and \cotwo\ snowlines, the majority of our models
have a C/H ratio close to what is expected based on snowline locations after 1~Myr of chemical evolution.
For these models the CO has undergone very little reprocessing and it remains in the gas at close to the initial abundance.
However, there is a subset of models where the gas phase C/H value is substantially reduced. These are the models with an enhanced cosmic ray ionization rate as well as the 0.003~\msun\ disk high X-ray models, i.e., those with significant CO reprocessing (see Figure~\ref{midheatin}).

After 6~Myr, chemical processes result in substantially lower gas phase C/H for most models.  
The models with some reduction in gas phase carbon after 1~Myr retain very little gas phase carbon after 6~Myr. However, the 0.03 and 0.1~\msun\ models with a fiducial cosmic ray ionization rate (1.6\ee{-19} s$^{-1}$) retain much of their initial CO gas even after 6~Myr of evolution. In these models the gas C/H ratio has changed by less than a factor of two after 6~Myr of chemical evolution.
In summary, models with greater CO reprocessing have a lower C/H ratio in the gas.

After 1~Myr 22\% of models have a C/H ratio less than half that predicted by \citep{Oberg11c}, increasing to 41\% after 3~Myr and 49\% at both 5 and 6~Myr. This suggests planets formed in young, a few~Myr old, disks are unlikely to have initial carbon abundances modified by disk chemistry unless there is a strong ionization source. 
Looking at the low CO gas abundance in the inner disk of TW Hya, chemistry alone is unlikely to be the sole cause, given the relatively large disk mass (0.05~\msun) and low cosmic ray ionization rate \citep{Cleeves15,Schwarz16}. Our models consider chemistry for a static disk model, without dynamics or an evolving dust population. Such mechanisms must also be at work in order to explain the low observed abundances. As it stands, without the presence of cosmic rays, disk chemistry is unlikely to have a substantial impact on the total gas phase carbon available to forming planets near 12~au. 
 
 \subsubsection{O/H}
In this section we analyze the total gas phase O/H ratio in the midplane at 12~au. To calculate the O/H based only on snowline locations we assume 34\% of the total oxygen is in CO, 14\% is in \cotwo, 20\% is in \water, and the remaining 32\% is in refractory silicates. Between the CO and \cotwo\ snowlines the predicted O/H ratio relative to stellar is 0.34, increasing to 0.48 inside the \cotwo\ snowline and 0.68 inside the \water\ snowline.

\begin{figure}
\setlength{\intextsep}{0pt}
\centering
    \includegraphics[width=0.5\textwidth]{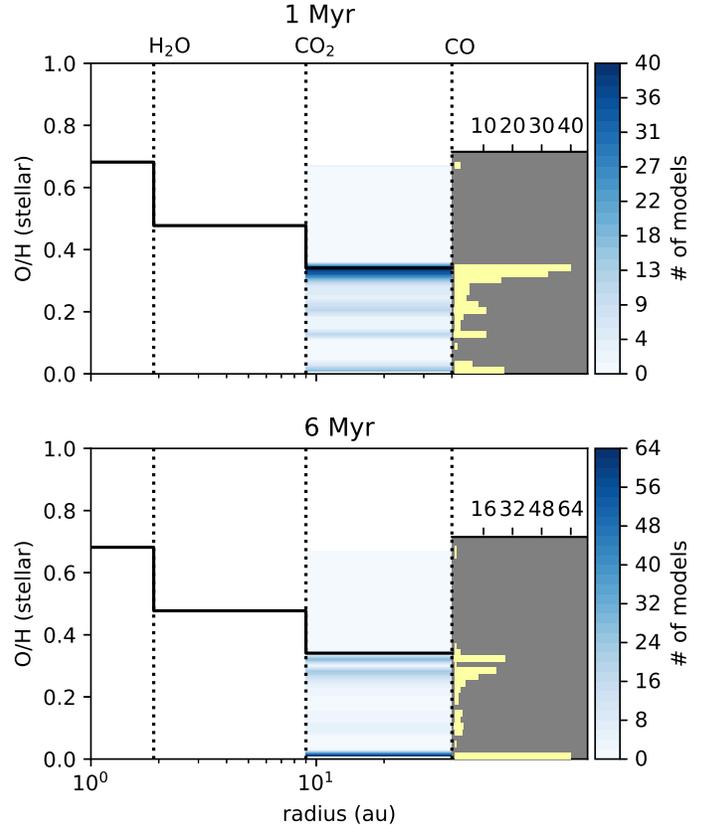}
\caption{O/H ratio in the gas based on the location of major volatile snowlines (line) and the distribution of O/H values inside the CO snowline for our models (shading \& histogram) after 1~Myr (top) and 6~Myr (bottom). Radial changes in the O/H value are based on the oxygen partitioning assumptions of \citet{Oberg11c}.
\label{ohplot}}
\end{figure}

Comparing this predicted O/H profile to the O/H gas abundances in our models inside the midplane CO snowline, the breakdown of models with different O/H ratios follows that seen for C/H (Figure~\ref{ohplot}). This is unsurprising since CO is the dominant gas phase reservoir for both carbon and oxygen. 
On the whole, the  spread of O/H values in our models is smaller than for C/H since most of the oxygen is in \water\ ice.
There are a handful of models with an extremely low O/H ratio, significantly lower than the corresponding C/H, in which there are large reservoirs of \methanol\ ice in addition to \cotwo\ ice and \water\ ice. This leads to an excess of carbon with respect to oxygen in the gas, the consequences of which are discussed in the next section.

\subsubsection{C/O}
\begin{figure}
\setlength{\intextsep}{0pt}
\centering
    \includegraphics[width=0.5\textwidth]{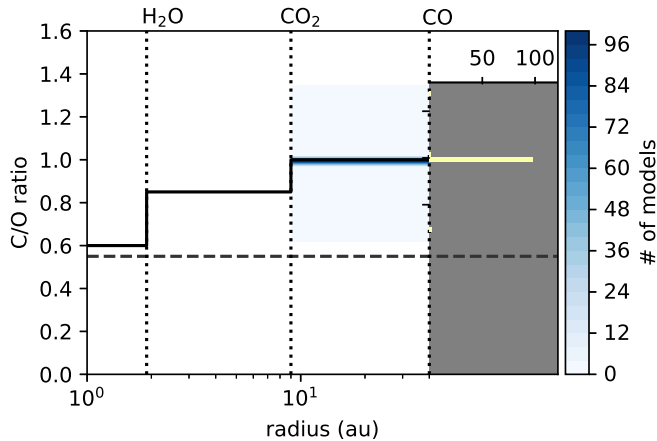}
\caption{C/O ratio in the gas based on the location of major volatile snowlines (line) and the distribution of C/O values inside the CO snowline for our models (shading \& histogram) after 6~Myr. Models with extremely large C/O are not shown. The grey dashed line indicates C/O for the Sun. \label{CoverO}}
\end{figure}

Figure~\ref{CoverO} shows the midplane C/O ratio in the gas at 12~au after 6~Myr. Nearly every model has C/O $\sim$ 1. Only one model, the warm, high cosmic ray rate model with a disk mass of 0.03~\msun\ and 99\% of dust in large grains, has a C/O ratio below solar.
Thus, any planet formed between the CO and \cotwo\ snowlines observed to have a sub-stellar atmospheric C/O would need to obtain at least half of its metals from solids. As this seems unlikely, the atmospheres of such planets should have super-stellar C/O.

\begin{figure}
\setlength{\intextsep}{0pt}
\centering
    \includegraphics[width=0.5\textwidth]{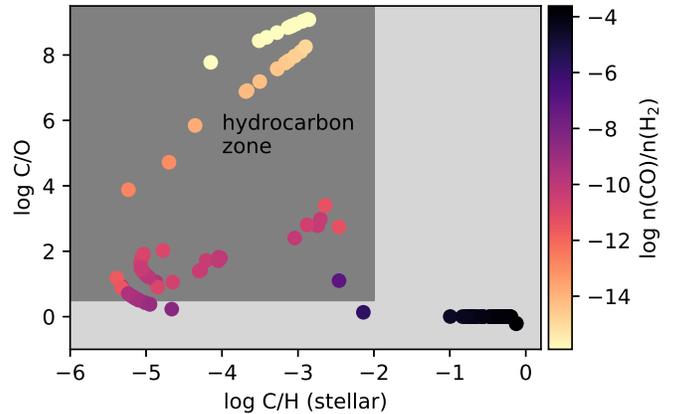}
\caption{Gas phase C/O vs. gas phase C/H relative to stellar for the midplane at 12~au after 6~Myr. When cosmic rays are present some of the carbon in CO is converted into gas phase hydrocarbons, resulting in a large C/O. \label{covch}}
\end{figure}

While most models have a midplane C/O ratio near unity, the high cosmic ray rate models with extreme CO reprocessing have extremely large C/O ratios of order \eten{8} (Figure~\ref{covch}). In these models both the gas phase carbon and oxygen are substantially lower.
While most of the carbon is in \methanol\ ice and \cotwo\ ice, the most abundant gas phase carbon species is CH$_4$ with abundances up to \eten{-7}. In these models C$_2$H is the second most abundant gas phase carbon species, though abundances only reach \eten{-11}.  

Within our time dependent chemical model, the series of chemical reactions leading to CH$_4$ begin when He$^+$ reacts with CO to create C$^{+}$, which recombines with an electron. The resulting neutral carbon atom reacts with \hh\ to form CH$_2$, which then freezes out. Successive hydrogenation on the grain surface forms CH$_4$ ice. While the majority of the CH$_4$ molecules remain on the grain surface, some do desorb back into the gas. This population of gas phase hydrocarbons is built up over time, and is not seen at earlier times when the total gas phase O/H is high.
In these models the gas phase abundances of both carbon and oxygen have been reduced, with the oxygen more depleted than the carbon. 
Similar reductions in the higher, UV-dominated regions of the disk likewise are needed to match the rings of C$_2$H emission observed in TW Hya and DM Tau \citep{Bergin16}.

The presence of gas phase hydrocarbons at a low abundance results in an extreme enhancement of the gas C/O ratio.
Sublimation of unprocessed CO ice from the outer disk as grains drift inward could easily dilute this effect. However, as can be seen in Figures~\ref{fullmod1}~\&~\ref{fullmod6}, much of the carbon in the outer disk is chemically reprocessed into less volatile ices, which will not sublimate near the CO snowline. Additional work is needed in order to understand how the combined effects of dust drift and chemistry change the gas phase C/O ratio in the inner disk.

\subsection{Comparison to Comets}

Comets, as the least modified bodies, provide the best record of the composition of volatile ices in the solar nebula. Surveys of primary volatiles in Oort cloud comets reveal high abundances of \cotwo, CO, and \methanol\ relative to water \citep{Mumma11}. 
Oort cloud comets formed much closer in, 5-15 au, but were scattered out of their formation region by an early disruption even, e.g., as predicted by the Nice Model \citep{Gomes05}

\begin{figure*}
\setlength{\intextsep}{0pt}
    \includegraphics[width=1.0\textwidth]{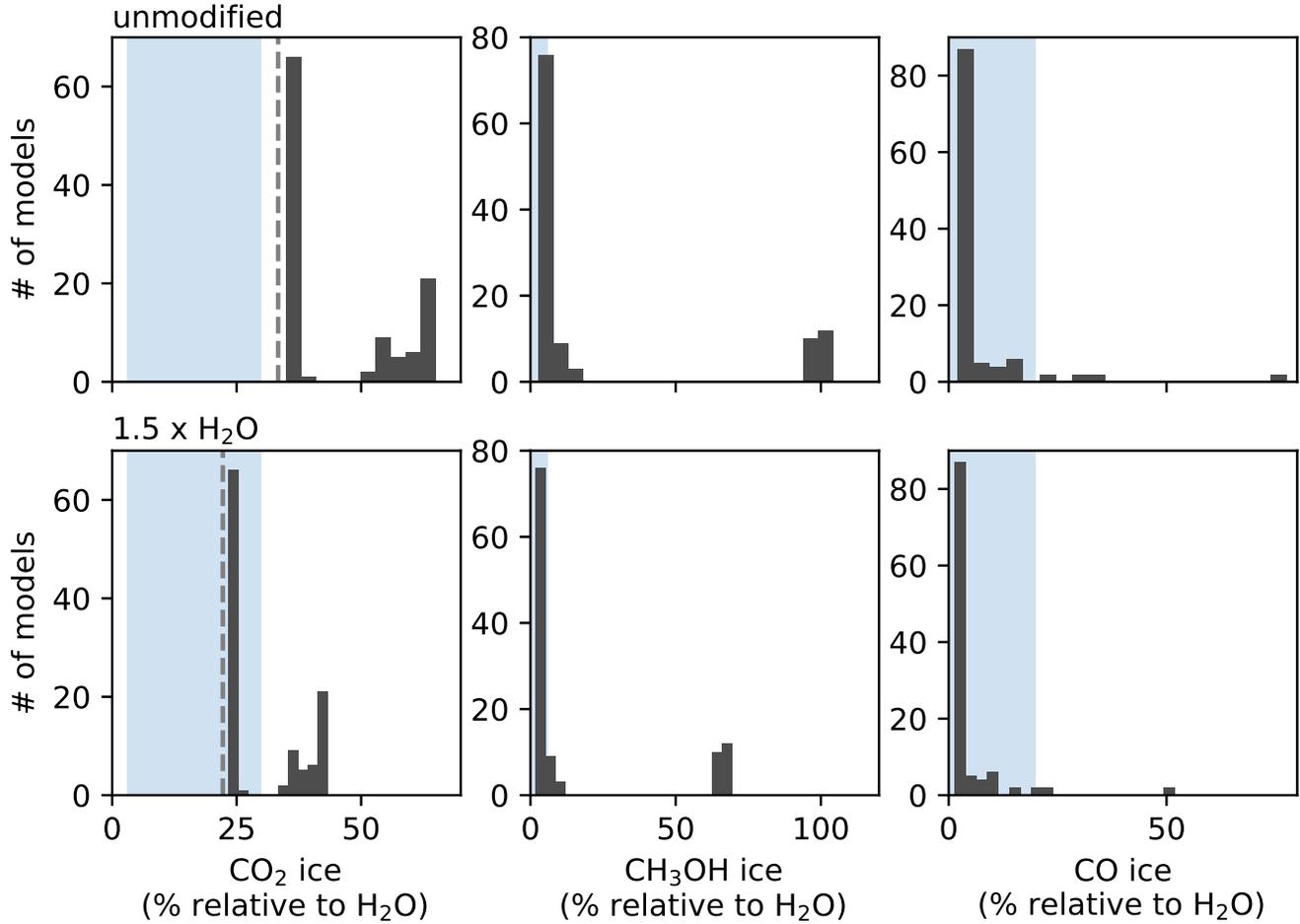}
\caption{Distribution of ice abundances in all of our 0.1 and 0.03~\msun\ models relative to \water\ ice after 1 Myr. Blue shaded regions show range of ice abundances observed in comets and grey dashed lines indicate the initial \cotwo\ abundance. Cometary values are taken from \citet{Mumma11}. \label{comets}}
\end{figure*}

Figure~\ref{comets} compares the midplane abundances of \cotwo, CO, and \methanol\ ices in our models to the range of values observed in comets. We consider our models with a total disk mass of either 0.1~\msun\ or 0.03~\msun, as the surface density at 12~au in these models bracket the value for the minimum mass solar nebula \citep{Weidenschilling77}. We do not include the models for which the initial \water\ abundance was removed, but do include all models with our fiducial initial abundances.
Our models already begin with a \cotwo\ ice abundance relative to water greater than that seen in comets, which only increases.
In contrast, our models do not start with any \methanol\ or CO ice, but the majority of models are able to match the abundance of these ices relative to \water\ after 1~Myr of chemical evolution. After 6~Myr most models contain more \methanol\ ice than seen in comets while still matching for CO ice.

Since 12~au is well outside the midplane \water\ snowline, the midplane \water\ ice abundance evolves very little with time. Therefore, we can easily explore the affect of a different initial \water\ abundance.
Increasing the \water\ ice abundance by a factor of 1.5 brings \cotwo\ ice abundances within the range observed in comets while also keeping CO and \methanol\ ices within the cometary range for the majority of the models. Alternatively, starting with less carbon in \cotwo\ ice would lead to \cotwo\ ice abundances within the range observed in comets after some chemical evolution. However, the CO ice and \methanol\ ice abundances in such models cannot be predicted without running the chemistry.
In summary, the ice abundances in our models are strongly sensitive to the initial conditions and do not provide stringent constraints on the origins of cometary ices. However, the effects seen in our models do demonstrate that the ice abundances in disks can reproduce those observed in comets for a range of physical condition.

\section{Summary}\label{midsummary}
We have analyzed the chemical abundances in the midplane at 12~au for 198 unique, physically static, disk models, considering a range of disk mass, large grain fraction, X-ray luminosity, cosmic ray ionization, temperature, and initial \water\ abundance with the goal of exploring the range of physical conditions under which carbon can be removed from CO via chemical reprocessing. We find that:
\begin{enumerate}
\item Under most conditions an ISM level cosmic ray ionization rate is needed to reprocess CO in the midplane, converting it to \methanol\ ice.

\item In highly dust settled, low surface density disks the presence of UV photons can result in CO being reprocessed into \cotwo\ ice. Thus disk gas evolution can lead to substantial chemical evolution in the main elemental pools of carbon and oxygen.

\item While most models have gas phase C/H and O/H abundances close to what is expected based on snowline locations, models with more CO reprocessing have lower C/H and O/H, with the most depleted models having extremely low O/H since the formation of \cotwo\ ice preferentially removes oxygen.

\item The gas phase C/O in most models is near unity, with only one model having a sub-solar C/O. Several models have highly elevated C/O values. These are models highly depleted in CO due to the presence of a strong source of ionization, but with a significant reservoir of gas phase hydrocarbons.

\item Our models over-produce  \cotwo\ ice relative to \water\ as compared to cometary abundances, but fall within the range observed for \methanol\ and CO ice. Modifying the initial ice abundances allows us to match cometary abundances for all three species in a subset of models.
\end{enumerate}

\vskip 0.1in

We conclude that a strong source of ionization or photolysis is needed to chemically reprocess CO in the disk midplane via in situ processes. While cosmic ray ionization is unlikely to contribute to the low midplane CO abundances in TW Hya given its low ionization rate, a more complete understanding of the ionization structure in protoplanetary disks as a population is needed in order the determine the contribution of chemistry to the removal of volatiles from the gas. Since it is extremely difficult for chemical processes to lower the gas phase C/O ratio between the CO and \cotwo\ snowlines, any planets accreting their atmospheres in this region should have super-stellar C/O in the absence of other processes. 

\acknowledgments
This work was supported by funding from NSF grant
AST-1514670 and NASA NNX16AB48G. 
K.S. and K.Z. acknowledge the support of NASA through Hubble Fellowship Program grants HST-HF2-51419.001 and HST-HF2-51401.001, awarded by the Space Telescope Science Institute, which is operated by the Association of Universities for Research in Astronomy, Inc., for NASA, under contract NAS5-26555.

\software{GNU Parallel \citep{gnuparallel}, IDL, matplotlib \citep{matplotlib}, numpy \citep{numpy}, pandas \citep{pandas}, scipy \citep{scipy}, TORUS \citep{Harries00}}

\appendix

\section{Truncated Disk}\label{app:smalldisk}
\begin{figure}[!]
\setlength{\intextsep}{0pt}
\centering
    \includegraphics[width=0.5\textwidth]{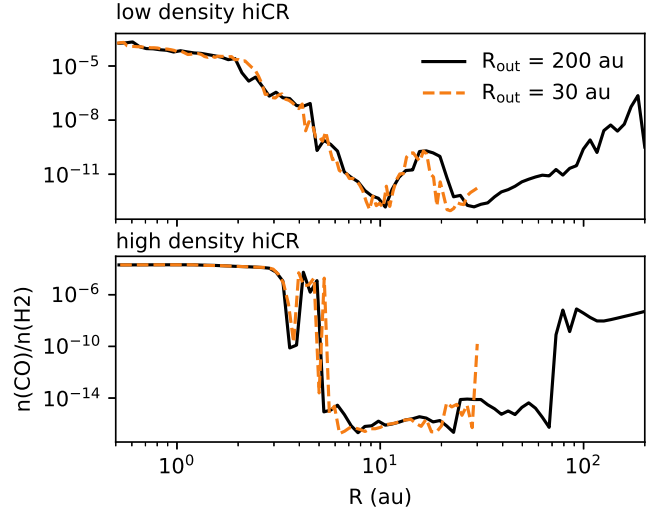}
  \caption{Comparison of midplane CO gas abundance abundance after 6 Myr of chemical revolution in disks with an outer radius of 200 au and 30 au. Top: Disks with a surface density equivalent to the surface density in our 0.003\msun\ models with an outer radius of 200~au, 50\% large grains and a high cosmic ray ionization rate. Bottom:  Disks with a surface density equivalent to the surface density in our 0.1\msun\ models with an outer radius of 200~au, 50\% large grains, and a high cosmic ray ionization rate.\label{fig:app:smallmods}}
\end{figure}

To test the sensitivity of the midplane chemistry to our choice of outer radius we generated a disk model with the same surface density as our 0.003 \msun\ and 0.1 \msun\ disks with 50\% in large grains, but with the outer radius truncated to 30 au. We then reran the full radiative transfer and chemistry for a cosmic ray ionization rate of 2\ee{-17} s$^{-1}$ (our high cosmic ray rate case) and an X-ray luminosity of \eten{30} erg s$^{-1}$ (our fiducial X-ray luminosity). Figure~\ref{fig:app:smallmods} compares the midplane CO gas abundances in our truncated disk to the equivalent 200~au disk. There is very little variation in the CO abundance as a function of disk size. In part, this is because the only source of radiation in our models is the central star. As such radiation does not need to pass through, and be attenuated by, the outer disk in order to reach the inner disk. Additionally, the chemistry at each radius is calculated independently such that the abundances at one radius do not affect abundances at a smaller radius via, e.g., self-shielding. Finally, we note that while our models are physically static, radial drift in an evolving disk could result in different chemical abundances between large and small disks.

\section{Abundance Table}\label{app:table}

Table~\ref{app:tablemidin} list the five most abundance carbon bearing species in the midplane at 12~au for each model.

%\begin{landscape}
\begin{deluxetable}{lrrrrrrrrrrrr}[!h]
\tabletypesize{\scriptsize}
\tablewidth{0pt}
\tablecaption{Top five most abundant carbon bearing species in the midplane at 12~au for each model after 1~Myr. Abundances are relative to \hh. Full table available online.}\label{app:tablemidin}
\tablehead{
\colhead{Model} & \colhead{M$\mathrm{_{disk}}$} & \colhead{$f_l$ } & \colhead{Species} & \colhead{Abundance} & \colhead{Species} & \colhead{Abundance} & \colhead{Species} & \colhead{Abundance} & \colhead{Species} & \colhead{Abundance} & \colhead{Species} & \colhead{Abundance}  \\
}
\startdata
fiducial             & 0.003  & 0.0   & CO         &    1.93e-04 & CO2(gr)    &                      4.24e-05 & HCN(gr)    &    7.76e-07 & HNC(gr)    &    7.34e-07 & CH3OH(gr)                    &    5.27e-07 \\ 
fiducial             & 0.003  & 0.1   & CO         &    1.94e-04 & CO2(gr)    &                      4.21e-05 & HCN(gr)    &    7.63e-07 & HNC(gr)    &    7.28e-07 & CH3OH(gr)                    &    4.97e-07 \\ 
fiducial             & 0.003  & 0.2   & CO         &    1.94e-04 & CO2(gr)    &                      4.22e-05 & HCN(gr)    &    7.46e-07 & HNC(gr)    &    7.15e-07 & CH3OH(gr)                    &    5.57e-07 \\ 
fiducial             & 0.003  & 0.3   & CO         &    1.92e-04 & CO2(gr)    &                      4.30e-05 & CH3OH(gr)  &    8.44e-07 & HCN(gr)    &    6.93e-07 & HNC(gr)                      &    6.70e-07 \\ 
fiducial             & 0.003  & 0.4   & CO         &    1.88e-04 & CO2(gr)    &                      4.40e-05 & CH3OH(gr)  &    1.25e-06 & HCN(gr)    &    6.10e-07 & HNC(gr)                      &    5.97e-07 \\ 
fiducial             & 0.003  & 0.5   & CO         &    1.88e-04 & CO2(gr)    &                      4.42e-05 & CH3OH(gr)  &    1.21e-06 & HCN(gr)    &    7.14e-07 & HNC(gr)                      &    6.70e-07 \\ 
fiducial             & 0.003  & 0.6   & CO         &    1.87e-04 & CO2(gr)    &                      4.45e-05 & CH3OH(gr)  &    1.48e-06 & C3H4(gr)   &    5.85e-07 & HCN(gr)                      &    5.43e-07 \\ 
fiducial             & 0.003  & 0.7   & CO         &    1.82e-04 & CO2(gr)    &                      4.60e-05 & CH3OH(gr)  &    2.30e-06 & C3H4(gr)   &    8.03e-07 & HC3N(gr)                     &    5.58e-07 \\ 
fiducial             & 0.003  & 0.8   & CO         &    1.79e-04 & CO2(gr)    &                      4.67e-05 & CH3OH(gr)  &    2.76e-06 & C3H4(gr)   &    9.25e-07 & HC3N(gr)                     &    7.53e-07 \\ 
fiducial             & 0.003  & 0.9   & CO         &    1.74e-04 & CO2(gr)    &                      4.66e-05 & CH3OH(gr)  &    3.91e-06 & HC3N(gr)   &    1.36e-06 & C3H4(gr)                     &    1.18e-06 \\ 
\enddata
\end{deluxetable}
%\end{landscape}

\end{document}